\documentclass{ws-procs975x65}

\usepackage{psfig}

\begin{document}

\title{Large energy dependence of current noise in
superconducting/normal metal junctions}

\author{F. Pistolesi}
\address{Laboratoire de Physique et Mod\'elisation des Milieux Condens\'es,\\
CNRS-UJF B.P. 166, F-38042 Grenoble, France}

\author{M. Houzet}
\address{ Commissariat \`{a} l'\'{E}nergie Atomique, DSM,\\
D\'{e}partement de Recherche Fondamentale sur la Mati\`{e}re
Condens\'{e}e, SPSMS, F-38054 Grenoble, France }

\newcommand{\qav}[1]{\left\langle #1 \right\rangle}
\newcommand{\myT}{\Gamma}
\newcommand{\rem}[1]{}
\newcommand{\FORSE}[1]{{\bf Peut-etre #1} }
\newcommand{\refe}[1]{(\ref{#1})}
\newcommand{\refE}[1]{Eq.~(\ref{#1})}
\newcommand{\beq}{\begin{equation}}
\newcommand{\eeq}{\end{equation}}
\newcommand{\beqa}{\begin{eqnarray}}
\newcommand{\eeqa}{\end{eqnarray}}
\newcommand{\cg}{\check g}
\newcommand{\inc}{{\rm inc}}

\newcommand{\Gc}{G}
\newcommand{\Fc}{F}
\newcommand{\Dc}{C}

\maketitle

\abstracts{
Interference of electronic waves undergoing Andreev reflection in
diffusive conductors determines the energy profile of the
conductance on the scale of the Thouless energy.
A similar dependence exists in the current noise, but its behavior
is known only in few limiting cases.
We consider a metallic diffusive wire connected to a
superconducting reservoir through an interface characterized by an
arbitrary distribution of channel transparencies.
Within the quasiclassical theory for current fluctuations we
provide a general expression for the energy dependence of the
current noise.
We derive closed analytical expressions for large energy.
}

\section*{Introduction}

Interference of electronic waves in metallic disordered conductors
is responsible for weak localization corrections to the
conductance.\cite{Weakloc}
If these are neglected,  the probability of transferring an
electron through the diffusive medium is given by the sum of the
{\em modulus squared} of the quantum probability amplitudes for
crossing the sample along all possible paths.
This probability is denoted as semiclassical, since quantum
mechanics is necessary only for establishing the probability for
following each path independently of the phases of the quantum
amplitudes.
In superconducting/normal metal hybrid structures, interference
contributions are not corrections, they may actually {\em
dominate} the above defined semiclassical result for temperatures
and voltages smaller than the superconducting gap.
This is seen experimentally as an energy dependence of the
conductance on the scale of the Thouless energy.
Indeed, the energy dependence comes from the small wavevector
mismatch, linear in the energy of the excitations, between the
electron and the Andreev reflected hole.
This is responsible for the phase difference in the amplitudes for
two different paths leading to interference.
The effect is well known and explicit predictions and measurements
exist for a number of systems.\cite{VZK93,Hekking93,Beenakker97}

Interference strongly affects the current noise too.\cite{BlanterButtiker}
The largest effects are expected in the tunneling limit, when the
transparency of the barrier is small and its resistance is much
larger than the resistance of the diffusive normal region.
Then, the conductance has a strong non linear dependence at low
bias (reflectionless tunneling).\cite{VZK93,Hekking93}
This is actually the case, but the zero-temperature noise (or shot
noise) does not give any additional information on the system
since it is simply proportional to the current, as shown quite
generally in Ref. \cite{PBH}.
In the more interesting case of a diffusive metal wire in contact
with a superconductor through an interface of conductance $G_B$
much larger than the wire conductance $G_D$, Belzig and Nazarov\cite{BN01}
found that the differential shot noise, $dS/dV$, shows
a reentrant behavior, as a function of the voltage bias, {\em
similar}, but not {\em identical}, to the conductance one.
(The extension of the Boltzman-Langevin approach to the coherent
regime in Ref.\ \cite{houzet} neglects this difference.)
In order to compare quantitatively with actual experiments\cite{Jehl,KSP,Lefloch}
and to gain more insight in the
interference phenomenon, it is necessary to obtain the energy
dependence of noise in more general situations.
The numerical method used in Ref.~\cite{BN01} is, in principle,
suitable to treat more general cases, notably the case when $G_D
\sim G_B $, but only if all channel transparencies, $\{\myT_n\}$,
that characterize the interface are small.
In Ref. \cite{HouzetPistolesi} we presented an analytical solution
for the diffusion-type differential equation for the noise within
the theory of current fluctuations\cite{FCS} in the
quasiclassical dirty limit.\cite{BN01}
It allows to treat the general case of arbitrary values for
$\{\myT_n\}$ and $G_B/G_D$.
In the present paper we present the minimal set of equations
necessary to obtain the noise. We then exploit them to obtain
closed analytical expressions for the noise at large energy.

\section{Equations to obtain the noise}

In Ref. \cite{HouzetPistolesi} we have developed an analytical
theory to calculate the current noise in a diffusive wire of
length $L$, diffusive constant $D$, conductance $G_D$, connected
to a normal reservoir on one side (with a transparent junction)
and to superconductor on the other side through an arbitrary
interface characterized by a set of channel transparencies
$\{\myT_n\}$.
These equations are obtained by exploiting the semiclassical
theory proposed by Nazarov\cite{CircuitTheory} to calculate the
full counting statistics of charge transfer\cite{FCS}.  Details
are given in Ref. \cite{HouzetPistolesi}.
Let us summarize in a compact form the equations necessary to
obtain the noise.

The first step is to obtain the conductance.
This depends on $f_T(x)$ and $\theta(x)$, parameterizing the
fermion distribution and the superconducting correlations,
respectively.
The variable $x$ varies between 0 and $L$, $\theta=\theta_1+i
\theta_2$ is a complex number and satisfies the following equation:
\beq \label{eqtheta}
   \hbar\,D \, \theta''(x)+2i\varepsilon\sinh\theta(x)=0
   \,,
\eeq
with boundary conditions\cite{zaitsev} $\theta(L)=0$ and $
\theta'(0)= H[\theta(0)]$ where
\beq \label{BCtheta}
    H[\theta]
    = {i\over L\,r}
    \qav{
    \cosh \theta
    \over 1+{\myT\over 2}
    \left(i \sinh \theta-1\right)
} \,. \eeq
We defined $
  \qav{\psi (\myT)}
  \equiv
  \sum_n \myT_n \psi(\myT_n)/\sum_n \myT_n
$ and $r=G_D/G_B$, with  $G_B= (2 e^2/h) \sum_n \myT_n$.
For $f_T$ we have the simpler equation
$(\cosh^2 \theta_1(x) f_T'(x))'=0$
with boundary conditions $f_T(L)=f_{T0}$ and
\beq \label{BCfT}
 f_T'(0)
 =
 {f_T(0)\, \theta'_1(0) \over \cosh \theta_1(0)\,
 \sinh \theta_1(0)}
 \quad.
\eeq
Here $f_{T0}=f_--f_+$, $f_{\pm}(\varepsilon)=f(\varepsilon\pm
eV)$, $f$ is the Fermi function at temperature $T$, and $V$ is the
voltage bias.
Then the current is
$
    I = 1/(2e)
    \int \!\! d\varepsilon \,\Gc(\varepsilon)
    f_{T0}(\varepsilon)
$
 with\cite{VZK93}
\beq \label{eqG}
    {\Gc} (\varepsilon)
    =
    G_D \left[
  {\Dc}^{-1}(\varepsilon)+
  {\tanh \theta_1(0)\over L\, \theta'_1(0)}
  \right]^{-1}
\eeq and $
  {\Dc}^{-1}(\varepsilon)
  =
  1/L \int_0^L ds/\cosh^{2}\theta_1(s). $
At low temperatures, $k_B T\ll eV$, $G_{diff}(V)={\Gc}(eV)$ is the
differential conductance.
These equations have been recently exploited in Ref.~\cite{tanaka}
to discuss the conductance.

To obtain the noise we need an additional parameter $a(x)$. It
parameterizes the first correction in the counting field to the
Usadel Green's function. Other parameters intervene, but we do not
need them to calculate the noise.\cite{HouzetPistolesi}

The complex parameter $a=a_1+i a_2$ satisfies the following linear
differential equation:
\beq
 \label{eqfora}
 \hbar\,D \,a''(x) + 2i\, \varepsilon\,  a(x) \cosh\theta(x)
 =
 -2 E_{T}\,
 {\sinh\theta_1(x) \over \cosh^3 \theta_1(x)}
 \frac{G(\varepsilon)^2}{G_D^2}
 \,,
\eeq
with $E_T=\hbar D/L^2$.
The boundary conditions are $a(L)=0$ and $L\,a'(0)=\alpha\, a(0)/r
+\beta/r$ with
\beqa \label{BCa} \alpha &=& \left\langle { i \sinh
\theta-\myT(i\sinh
    \theta-1)/2 \over \left[1+\myT(i\sinh \theta-1)/2\right]^2 }
\right\rangle \\
\beta &=& {i c^2 \over 8} \left\langle {2\myT^2\cosh
\theta^*+8(\myT-1) \cosh\theta -2 i\myT(\myT-2)\sinh \theta \cosh
\theta^* \over \left|1+\myT(i\sinh \theta-1)/2\right|^2
\left(1+\myT(i\sinh \theta-1)/2\right) } \right\rangle \,, \eeqa
%
both evaluated at $x=0$ and we defined $r=G_D/G_B$.
The low frequency noise is finally given by:
\begin{equation} \label{eq:bruit}
  S
  =
  \int \!\! d\varepsilon \, {\Gc}(\varepsilon)
  \left\{ 1-f_{L0}^2(\varepsilon)-[1-{\Fc}(\varepsilon)]f_{T0}^2(\varepsilon) \right\},
\end{equation}
where
\begin{eqnarray}
\lefteqn{ {\Fc}(\varepsilon)=\frac{2}{3}(1+c(0)^3) + {2{\Gc}(\varepsilon)\over G_D} \int_0^1 {\sinh \theta_1 a_1 \over
\cosh^3 \theta_1}ds}
\nonumber \\
&& -c(0) \left(
     {G_D a_1'(0) c(0) \over {\Gc}(\varepsilon) \tanh \theta_1(0)}
     +{2 a_1(0) \over \sinh 2\theta_1(0)}
\right) \,. \label{eq:fano}
\end{eqnarray}
Here the parameter $c$ is simply poportional to $f_T(x)$:
$c(x)=-f_T(x)/f_{T0}$ with $c(0)=1-{\Gc}(\varepsilon)/[G_D{\Dc}(\varepsilon)]$, and $f_{L0}=1-f_+-f_-$.

We have all ingredients to calculate explicitly the noise for
arbitrary values of the ratio $r$, of the energy $\varepsilon$, or
of the transparency set $\myT_n$.

\section{Large energy limit for $r\neq 0$}

In the following we discuss the $\varepsilon \gg E_T $ analytical
limit. If $r \neq 0$ for large $\varepsilon$ the parameter
$\theta$ is very small, it is thus possible to set up the
following expansion:
\beq \label{expansion}
\theta(x) = {\theta^{(0)}(x) \over k } +
{\theta^{(1)}(x) \over k^2 } + {\theta^{(2)}(x) \over k^3 } +
\dots \eeq
where we introduced the large parameter of the expansion
$k=\sqrt{\varepsilon}$. We can now substitute \refe{expansion}
into \refe{eqtheta} and into the boundary conditions. In
collecting terms of the same order in $1/k$ one has to remember
that each derivative with respect to $x$ introduce a $k$ factor.
At lowest order we obtain
\beqa
  && D\, {\theta^{(m)}}''(x) + 2 i k^2\,\theta^{(m)} = 0  \label{diff1}\\
  && D\,{\theta^{(2)}}''(x) + 2 i k^2\,[\theta^{(2)} +{\theta^{(0)}}^3] = 0
    \label{diff2}
  \, .
\eeqa where $m=0,1$.
For the boundary conditions we obtain $\theta^{(m)}(L)=0$ for all
$m$, $\theta'^{(0)}=k \, H(0)$, $\theta'^{(1)}=k \, H'(0)
\theta^{(0)}$, and $\theta'^{(2)}=k \, H'(0) \theta^{(1)}+k \,
H''(0) {\theta^{(0)}}^2/2$.
To obtain the conductance to order $1/k^2$ we need also the
boundary condition for ${\theta^{(3)}}'$, but we do not need to
solve the associated differential equation.
Solving the differential equations \refe{diff1} and \refe{diff2}
and substituting the result into \refE{eqG} we obtain for the
conductance up to second order in $1/k$ the following expression:
\beq \label{Gana}
G = {H' \over  1+H'} + {H' H'' \over 2k(1+H')^2}
+ {H' H''(2H'+2{H'}^2-H H'') \over 4k^2(1+H')^4}+\dots \, .\eeq
Equation \refe{Gana} holds for any distribution of channel
transparency, it suffices to calculate the appropriate average in
the definition of $H$, $H'$, and $H''$.

The procedure to obtain the noise is similar. This time we need an
expansion of the parameter $a$ which has the same form of
\refe{expansion}. Actually the differential equation for $a^{(0)}$
and $a^{(1)}$ coincide with those for $\theta^{(0)}$. The equation
for $a^{(2)}$ reads:
\beq
   \hbar D \, {a^{(2)}}'' + 2 i k^2
   a^{(2)}= -2 k^2[i a^{(0)} {\theta^{(0)}}^2
   + f_{T0} G^2 \theta_1^{(0)}]\,.
\eeq
The boundary conditions for $a$ read: $a^{(0)}+k \beta^{(0)}/r=0$,
$a^{(1)}+k[\alpha{(0)}a^{(0)} +\beta^{(1)}]/r=0$,
$a^{(2)}+k[\alpha{(0)}a^{(1)} + \alpha{(1)}a^{(0)}+
\beta^{(2)}]/r=0$, where the same kind of development has been
performed on $\alpha$ and $\beta$. Substituting these expressions
into \refe{eq:fano} we can obtain the differential Fano factor
${\Fc} (\varepsilon)= (dS/ dV)(\varepsilon)/2e G(\varepsilon)$.
The lowest order, i.e. the incoherent contribution has a simple
expression: \beq {\Fc}_{inc} = \frac{2}{3} \left[ 1+
(2-3\frac{\left\langle\frac{\myT^3}{(2-\myT)^4}\right\rangle}
{\left\langle\frac{\myT}{(2-\myT)^2}\right\rangle})\frac{G_D^3}{(G_D+2G_B\left\langle
\frac{\myT}{(2-\myT)^2} \right\rangle)^3} \,.\right] \label{Finc}
\eeq
This can be understood by a comparison with the classical
calculation of Ref. \cite{deJongBen} for a wire connected to
normal reservoirs. \refE{Finc} coincide with the Fano factor given
there when the substitution $e\rightarrow 2e$, $G_D\rightarrow
G_D/2$ and $\myT_n \rightarrow \myT_n^A=\myT_n^2/(2-\myT_n)^2$.
This is consistent with the expectation that phase coherence
becomes irrelevant at high energy (see also the discussion in Ref.
\cite{SB}).

The expression for the quantum correction to \refE{Finc} is
cumbersome in the general case and we will not present it.
Simpler expression is obtained when all transparencies are the
same ($\myT_n=\myT$):
\beq {\Fc}^{(1)} = {4 r^2(\myT-2)^2 \over  [ r(\myT-2)+2
\myT]^4 } \left[ -64 +\myT(\myT-2)(-64+r(\myT-2)(\myT^2+12
\myT-12)\right] \eeq
and ${\Fc}(\varepsilon)= {\Fc}_{inc}+{\Fc}^{(1)}/\sqrt{\varepsilon}$.

\section{Conclusions}

We presented a theory to calculate the energy dependence of the
noise in a wire connecting a normal with a superconducting
reservoir.
The theory allows to obtain closed analytical expressions in
different relevant limits. We considered here in some details the
large energy case. The classical incoherent result appears for
energy much larger than the Thouless energy. Quantum corrections
are explicitly evaluated when all transparencies have the same
value.

\section*{Acknowledgments}
F.P. acknowledge financial support
from CNRS/ATIP-JC 2002.
M.H. acknowledges financial support from the
ACI-JC no 2036 from the French Ministry of Research.

\end{document}